\documentstyle[12pt,epsfig]{article}
 \newcommand{\be}{\begin{eqnarray}}
 \newcommand{\ee}{\end{eqnarray}}
 \newcommand{\beq}{\begin{equation}}
 \newcommand{\eeq}{\end{equation}}
 \newcommand{\ba}{\begin{array}{1}}
 \newcommand{\ea}{\end{array}}
 \newcommand{\bb}{\begin{thebibliography}}
 \newcommand{\eb}{\end{thebibliography}}

 \newcommand{\abstitle}[1]{{\small {\bf #1}}}
 \newcommand{\absauthor}[1]{{\small {\bf #1}}}
 \newcommand{\address}[1]{{\it #1}}
 
\begin{document}
 \begin{center}
 \abstitle{PROTON STRUCTURE FROM HARD P-P PROCESSES AT HIGH ENERGIES}\\
 \vspace{0.6cm}
 \absauthor{G. I. Lykasov$^1$, A. A. Grinyuk$^1$, I. V. Bednyakov$^1$, Yu.Yu. Stepanenko$^{1,2}$ } 
\\ [0.6cm]
 \address{\it $^1$ Joint Institute for Nuclear Research -
  Dubna 141980, Moscow region, Russia\\
  $^2$ Gomel State University, Gomel 246019, Republic of Belarus 
}
 \end{center}
 \vspace{0.1cm}
 \vspace{0.2cm} 
\begin{center}
{\bf Abstract}
\end{center}
 \vspace{0.1cm}
     Up to now, the existence of  intrinsic (or valence-like) heavy quark 
     component of the  proton distribution functions has not yet been confirmed or rejected. 
     We show that this hypothesis can be verified at experiments on the inclusive production
     of the open strangennes (NA61) and at measurements of prompt photons or vector bosons 
     accompanied by heavy flavour jets performed at LHC, CERN.  Our theoretical study demonstrates
     that investigations of the intrinsic heavy quark contributions look very promising in hard 
     processes like $pp\rightarrow K^\pm+X$ and  $pp\rightarrow\gamma/Z/W +c(b)+X$.
     A possible observation of these components at the CBM, NICA experiments is discussed also.

\section{Intrinsic heavy flavours in the proton}
\label{I}
      The LHC, NA61 (CERN), CBM (Darmstadt) and NICA (Dubna) experiments can be a useful laboratory for 
      investigation of the unique structure of the proton, in particular for 
      the study of the parton distribution functions (PDFs) with high accuracy.
      It is well known that the precise knowledge of these PDFs is very important 
      for verification of the Standard Model and search for New Physics. 

      By definition, the PDF $f_a(x,\mu)$ is a function of the proton momentum fraction $x$ 
      carried by parton $a$ (quark $q$ or gluon $g$) at the QCD momentum transfer scale $\mu$. 
      For small values of $\mu$, corresponding to the long distance scales less than $1/\mu_0$, 
      the PDF cannot be calculated from the first principles of QCD 
      (although some progress in this direction 
      has been recently achieved within the lattice methods 
\cite{LATTICE}). 
      The PDF $f_a(x,\mu)$ at $\mu>\mu_0$ can be calculated by 
      solving the perturbative QCD evolution equations (DGLAP) 
\cite{DGLAP}.  
      The unknown (input for the evolution) functions $f_a(x,\mu_0)$ 
      can usually be found empirically from some 
      ``QCD global analysis'' 
\cite{QCD_anal1,QCD_anal2} of a large variety of data, typically at $\mu>\mu_0$. 

     In general, almost all $pp$ processes that took place at the LHC energies, 
     including the Higgs boson production,
     are sensitive to the charm $f_c(x,\mu)$ or bottom $f_b(x,\mu)$ PDFs. 
     Nevertheless, within the global analysis 
     the charm content of the proton at $\mu\sim\mu_c$ and 
     the bottom one at $\mu\sim\mu_b$ are both assumed to be negligible.
     Here $\mu_c$ and $\mu_b$ are typical energy scales relevant to the 
     $c$- and $b$-quark 
     QCD excitation in the proton.
     These heavy quark components arise in the proton only perturbatively
     with increasing $Q^2$-scale 
     through the gluon splitting in the DGLAP $Q^2$ evolution 
\cite{DGLAP}. 
     Direct measurement of the open charm and open bottom production 
     in the deep inelastic processes (DIS) confirms the perturbative 
     origin of heavy quark flavours 
\cite{H1:2005}. 
     However, the description of these experimental data is 
     not sensitive to the heavy quark distributions at relatively 
     large $x$ ($x>0.1$). 

\begin{figure}[h]
\centerline{\includegraphics[width=0.50\textwidth]{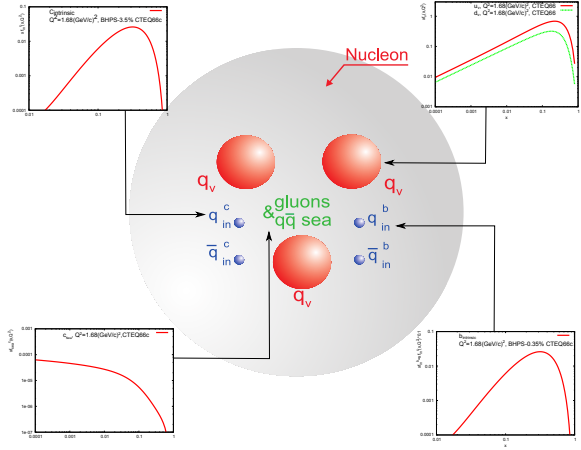}}
\caption{Schematic presentation of a nucleon consisting of three valence quarks
         q$_{\rm v}$, quark-antiquark q${\bar {\rm q}}$ and gluon sea, 
	 and pairs of the intrinsic charm 
	 (q$_{\rm in}^{\rm c}{\bar {\rm q}_{\rm in}^{\rm c}}$) 
	 and intrinsic bottom quarks 
	 (q$_{\rm in}^{\rm b}{\bar {\rm q}_{\rm in}^{\rm b}}$). Right top corresponds to valence
         $u(x)$ and $d(x)$ distributions at $Q^2=$1.68 (GeV$/$c)$^2$; the left bottom is the sea
         charm quark distribution; the left top is the {\it intrinsic} charm $x$-distribution
         and the right bottom is the  {\it intrinsic} bottom $x$-distribution at the same $Q^2$
         value.    
}
\label{Fig_IQ}
\end{figure}

     As was assumed by Brodsky with coauthors in 
\cite{Brodsky:1980pb, Brodsky:1981}, 
     there are {\it extrinsic} and {\it intrinsic} 
     contributions to the quark-gluon structure of the proton. 
     {\it Extrinsic} (or ordinary) quarks and gluons are generated on 
     a short time scale associated with a large-transverse-momentum processes.
     Their distribution functions satisfy the standard QCD evolution equations. 
     {\it Intrinsic} quarks and gluons exist
     over a time scale which is independent of any probe momentum transfer. 
     They can be associated with bound-state 
{(zero-momentum transfer regime)} hadron dynamics and 
     are believed to be of nonperturarbative origin.
Figure~\ref{Fig_IQ}  gives 
     a schematic view 
     of a nucleon, which consists of three valence quarks q$_{\rm v}$, 
     quark-antiquark q${\bar {\rm q}}$ and gluon sea, and, for example,  
     pairs of the {\it intrinsic} charm 
(q$_{\rm in}^{\rm c}{\bar {\rm q}_{\rm in}^{\rm c}}$) and 
     {\it intrinsic} bottom quarks 
(q$_{\rm in}^{\rm b}{\bar {\rm q}_{\rm in}^{\rm b}}$). 

       As is seen from Fig.~\ref{Fig_IQ}, the {\it intrinsic} charm and bottom distributions
       have the $x$-dependence similar to the valence quark ones (compare the left top to the right top),
       however the maximum of the  {\it intrinsic} charm magnitude is less than the valence one by a factor of
       10 and more if the {\it intrinsic} charm probability in proton is about 3.5$\%$. The  {\it intrinsic} bottom 
       distribution is similar to the {\it intrinsic} charm one, however its magnitude is less than the magnitude of
       {\it intrinsic} charm by factor 10, as it will be shown \cite{Polyakov:1998rb}. The distribution of the
       {\it extrinsic} (conventional perturbative) sea charm quarks in proton at $Q^2=$1.68 (GeV$/$c)$^2$ is presented 
       in the left bottom of Fig.~\ref{Fig_IQ}, it is suppressed a lot (about a few order) comparing to the {\it intrinsic} 
       charm distribution  at $x>$ 0.1.            

       It was shown in 
\cite{Brodsky:1981}
       that the existence of {\it intrinsic} heavy quark pairs 
       $c{\bar c}$ and $b{\bar b}$ within the proton state 
       could be due to the virtue of gluon-exchange and vacuum-polarization graphs. 
       On this basis, 
       within the MIT bag model 
\cite{Golowich:1981}, 
       the probability to find the five-quark component 
       $|uudc{\bar c}\rangle$ bound within the nucleon bag 
       was estimated to be about 1--2\%. 

       Initially in 
\cite{ Brodsky:1980pb,Brodsky:1981}
       S.Brodsky with coauthors 
       have proposed 
       existence of the 5-quark state $|uudc{\bar c}\rangle$ 
       in the proton 
(Fig.~\ref{Fig_IQ}). 
       Later some other models were developed. 
       One of them considered a quasi-two-body state 
       ${\bar D}^0(u{\bar c})\, {\bar\Lambda}_c^+(udc)$ in the proton 
\cite{Pumplin:2005yf}. 
       In 
\cite{Pumplin:2005yf}--\cite{Nadolsky:2008zw} 
       the probability to find the intrinsic charm (IC) in the proton 
       (the weight of the relevant Fock state in the proton)
       was assumed to be 1--3.5\%. 
       The probability of the intrinsic bottom (IB) in the proton 
       is suppressed by the factor $m^2_c/m^2_b\simeq 0.1$ 
\cite{Polyakov:1998rb}, where $m_c$ and $m_b$ are the masses of 
       the charmed and bottom quarks. 
       Nevertheless, it was 
       shown that the IC 
       could result in a sizable contribution 
       to the forward charmed meson production
\cite{Goncalves:2008sw}. 
       Furthermore the IC ``signal'' can 
       constitute almost 
       100\% 
       of the inclusive spectrum of $D$-mesons produced at 
       high pseudorapidities $\eta$ 
       and large transverse momenta $p_T$
       in $pp$ collisions at LHC energies 
\cite{LBPZ:2012}. 

       If the distributions of the intrinsic charm or bottom in the 
       proton are hard enough and are similar in the shape to the valence quark distributions
       (have the valence-like form),  
       then the production of the charmed (bottom) mesons or charmed (bottom) 
       baryons in the fragmentation region should be similar 
       to the production of pions or nucleons. 
       However, the yield of this production depends on the probability to find the 
       intrinsic charm or bottom in the proton, but this yield looks too small.     
       The PDF which included the IC contribution in the proton 
       have already been used in the perturbative QCD calculations in  
\cite{Pumplin:2005yf}-\cite{Nadolsky:2008zw}.

\begin{figure}[h]
\centerline{\includegraphics[width=0.50\textwidth]{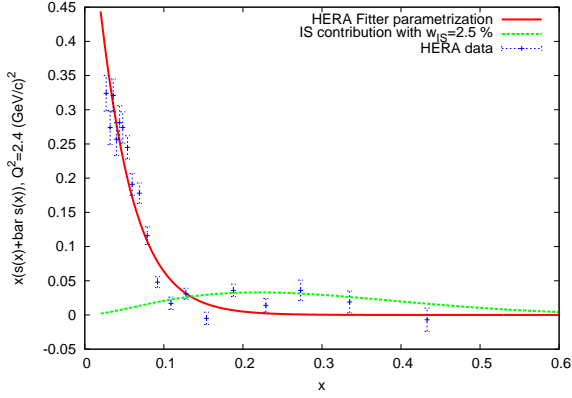}}
 \caption{The distributions of strange quarks $xS(x)=x(s(x)+{\bar s}(x))$ in the proton; the
solid line is the HERA Fitter parameterization of of $xS(x)$ at $Q^2=$2.4 GeV$/$c, the dashed curve 
is the contribution of the {\it intrinsic} strangeness (IS) in the proton with the probability 2.5 \%
}. 
\label{Fig_2IS}
\end{figure}

   Due to the nonperturbative {\it intrinsic} heavy quark components one can expect 
   some excess of the heavy quark PDFs over 
   the ordinary sea quark PDFs at $x>0.1$. 
   The ``signal'' of these components can be visible in the observables of 
   the heavy flavour production in semi-inclusive $ep$ DIS and inclusive 
   $pp$ collisions at high energies. 
   For example, it was recently shown that rather good description of the HERMES data 
   on the $xf_s(x,Q^2)+xf_{\bar s}(x,Q^2)$ at $x>0.1$ and 
   $Q^2=2.5$~GeV$/c^2$
\cite{IS:2012} could be achieved due to existence of {\it intrinsic} strangeness in the proton, 
   see Fig.~\ref{Fig_2IS}. One can see from Fig.~\ref{Fig_2IS} that the inclusion of the {\it intrinsic}
   strangeness allows us to describe the HERA data rather satisfactorily in the whole $x$-region both at
   $x\leq$ 0.1 and $x>$ 0.1.     

   Similarly, possible existence of the intrinsic charm in the proton
   can lead to some enhancement in the inclusive spectra of the open charm hadrons, 
   in particular $D$-mesons, produced at the LHC in $pp$-collisions 
   at high pseudorapidities $\eta$
   and large transverse momenta $p_T$ 
\cite{LBPZ:2012}.

       The probability distribution for the 5-quark state ($|uudc{\bar c}\rangle$) 
       in the light-cone description of the proton was first calculated in 
\cite{Brodsky:1980pb}. 
       The general form for this distribution calculated within the light-cone dynamics
       in the so-called BHPS model 
\cite{Brodsky:1980pb,Brodsky:1981} 
      can be written as 
\cite{IS:2012} 
\begin{eqnarray}
P(x_1,..,x_5)=N_5\delta\left(1-\sum_{j=1}^5x_j\right)
\left(m_p^2-\sum_{j=1}^5\frac{m_j^2}{x_j}\right)^{-2},
\label{def:B}
\end{eqnarray}
       where $x_j$ is the momentum fraction of the parton, $m_j$ is its mass and $m_p$ is the
       proton mass.  
       Neglecting the light quark ($u,d, s$) masses and the proton
       mass in comparison to the $c$-quark mass and integrating 
(\ref{def:B}) 
       over $dx_1...dx_4$ one can get the probability to find the  
       intrinsic charm with momentum fraction $x_5$ in the proton \cite{Peng_Chang:2012}: 
\begin{eqnarray}
P(x_5) &=& \frac{1}{2}{\tilde N}_5x_5^2 \Big[ \frac{1}{3}(1-x_5)(1+10x_5+x_5^2)- \nonumber \\ 
      && 2x_5(1+x_5)\ln(x_5) \Big],
\label{def:fcPumpl}
\end{eqnarray}
     where ${\tilde N}_5=N_5/m^4_{4,5}, m_{4,5}=m_c=m_{\bar c}$, 
     the normalization constant $N_5$ determines some 
     probability $w^{}_{\rm IC}$ to find 
     the Fock state $|uudc{\bar c}\rangle$ in the proton.

     As a rule, the gluons and sea quarks play the key 
     role in hard processes of open charm hadroproduction. 
     Simultaneously, due to the nonperturbative {\it intrinsic} heavy quark components 
     one can expect some excess of these heavy quark PDFs over 
     the ordinary sea quark PDFs at $x>0.1$. 
     Therefore the existence of this intrinsic charm component 
     can lead to some enhancement in the inclusive spectra of open charm hadrons, 
     in particular $D$-mesons, produced at the LHC in $pp$-collisions 
     at large pseudorapidities $\eta$ and large transverse momenta $p_T$ 
\cite{LBPZ:2012}.
    Furthermore, as we know from 
\cite{Brodsky:1980pb}-\cite{Nadolsky:2008zw}
    photons produced in association with heavy quarks $Q(\equiv c,b)$ in the final 
    state of $pp$-collisions provide valuable information about the parton 
    distributions in the proton
\cite{Pumplin:2005yf}-\cite{Thomas:1997}.

\smallskip
    In this paper,
    having in mind these considerations 
    we will first discuss where the above-mentioned 
    heavy flavour Fock states in the proton could be searched for 
    at the LHC energies. 
    Following this 
    we 
    analyze in detail, and give predictions for, the LHC 
    semi-inclusive $pp$-production of 
    prompt photons accompanied by  $c$-jets including the {\it intrinsic} 
    charm component in the PDF.


\section{Intrinsic heavy quarks in hard $pp$ collisions}
\label{II}
$\bullet~${\bf Where can one look for the intrinsic heavy quarks? }\\
    It is known that 
    in the open charm/beauty $pp$-production at large momentum transfer 
    the hard QCD interactions of two sea quarks, two gluons and 
    a gluon with a sea quark play the main role.
According to the model of hard scattering 
\cite{AVEF:1974}--\cite{FF:AKK08}
the relativistic invariant
inclusive spectrum of the hard process $p+p\rightarrow h+X$ can be related to
the elastic parton-parton subprocess $i+j\rightarrow i^\prime +j^\prime$,
where $i,j$ are the partons (quarks and gluons). 
This spectrum can be presented in the following general form 
\cite{FF}--\cite{FFF2} (see also \cite{BGLP:2011,BGLP:2012}):
\begin{eqnarray}
E\frac{d\sigma}{d^3p}= 
\sum_{i,j}\!\int\! d^2k_{iT}\!\int\! d^2k_{jT}\!\int_{x_i^{\min}}^1dx_i\!\int_{x_j^{\min}}^1dx_j\\
\nonumber
f_i(x_i,k_{iT})f_j(x_j,k_{jT}) 
\frac{d\sigma_{ij}({\hat s},{\hat t})}{d{\hat t}}\frac{D_{i,j}^h(z_h)}{\pi z_h}.
\label{def:rho_c}
\end{eqnarray}
   Here $k_{i,j}$ and $k_{i,j}^\prime$ are the four-momenta of the partons $i$ or $j$ 
   before and after the elastic parton-parton scattering, respectively; 
   $k_{iT}, k_{jT}$ are the transverse momenta of the partons $i$ and $j$;  
   $z$ is the fraction of the hadron momentum from the parton momentum; 
   $f_{i,j}$ is the PDF; and $D_{i,j}$ is the fragmentation function (FF) 
   of the parton $i$ or $j$ into a hadron $h$.

    When 
    the transverse momenta of the partons are neglected 
    in comparison with the longitudinal momenta, 
    the variables ${\hat s}$, ${\hat t}$, ${\hat u}$ and $z_h$ can be 
    presented in the following forms \cite{FF}:
\begin{eqnarray}
{\hat s}=x_i x_j s,\quad {\hat t}=x_i \frac{t}{z_h}, \quad
{\hat u}=x_j \frac{u}{z_h},\quad z_h=\frac{x_1}{x_i}+\frac{x_2}{x_j},
\label{def:stuzh}
\end{eqnarray}
     where
\begin{eqnarray}
x_1=-\frac{u}{s}=\frac{x_T}{2}\cot({\theta}/{2}), \quad
x_2=-\frac{t}{s}=\frac{x_T}{2}\tan({\theta}/{2}), \\
\nonumber
x_T=2\sqrt{t u}/s=2p_T/\sqrt{s}.
\end{eqnarray}
      Here as usual, 
      $s=(p_1+p_2)^2$,
      $t=(p_1-p_1^\prime)^2$,
      $u=(p_2-p_1^\prime)^2$, 
     and $p_1$, $p_2$, $p_1^\prime$ are the 4-momenta of the colliding protons 
     and the produced hadron $h$, respectively; 
     $\theta$ is the scattering angle for the hadron $h$ in the $pp$ c.m.s.
     The lower limits of the integration in
(\ref{def:rho_c}) are 
\begin{eqnarray}
x_i^{\min}=\frac{x_T \cot(\frac{\theta}{2})}{2-x_T \tan(\frac{\theta}{2})}, \qquad
x_j^{\min}=\frac{x_i x_T \tan(\frac{\theta}{2})}{2x_i-x_T \cot(\frac{\theta}{2})}.
\label{def:xijmn}
\end{eqnarray} 
Actually, the parton distribution functions $f_i(x_i,k_{iT})$ also depend on  
the four-momentum transfer squared $Q^2$ that is related to the Mandelstam variables
${\hat s},{\hat t},{\hat u}$ for the elastic parton-parton scattering \cite{FFF2} 
\begin{eqnarray}
Q^2~=~\frac{2{\hat s}{\hat t}{\hat u}}{{\hat s}^2+{\hat t}^2+{\hat u}^2}
\label{def:Qsqr}
\end{eqnarray} 

    Calculating spectra by Eq.(\ref{def:rho_c}) we used 
    the PDF which includes the IS (and does not include it) \cite{Nadolsky:2008zw},
     the FF of the type AKK08 
\cite{FF:AKK08}  and 
    $d\sigma_{ij}({\hat s},{\hat t})/d{\hat t}$ 
    calculated within the LO QCD and presented, for example, in 
\cite{Mangano:2010}.  

     One can see that the Feynman variable $x_F$ of the produced hadron, 
     for example, the $K$-meson, can be expressed via  
     the variables $p_T$ and $\eta$, or $\theta$ 
     the hadron scattering angle in the $pp$ c.m.s, 
\begin{eqnarray}
x_F \equiv \frac{2p_{z}}{\sqrt{s}}
=\frac{2p_T}{\sqrt{s}}\frac{1}{\tan\theta}
=\frac{2p_T}{\sqrt{s}}\sinh(\eta). 
\label{def:xFptteta}
\end{eqnarray} 
     At small scattering angles of the produced hadron this formula becomes 
\begin{eqnarray}
x_F\sim \frac{2p_T}{\sqrt{s}}\frac{1}{\theta}.
\label{def:xFtetapt}
\end{eqnarray} 
       It is clear that for fixed $p_T$ an outgoing hadron must possess a very small $\theta$ or very large $\eta$
        in order to have large $x_F$ (to follow forward, or backward direction).

       In the fragmentation region (of large $x_F$) the Feynman variable $x_F$ 
       of the produced hadron is related to 
       the variable $x$ of the intrinsic charm quark in the proton, and  
       according to the longitudinal momentum conservation law, 
       the $x_F \simeq x$ (and $x_F < x$). 
       Therefore, the visible excess of the inclusive spectrum, for example, of $K$-mesons
       can be due to the enhancement of the IS distribution 
(see Fig.~\ref{Fig_2IS}) at $x>$ 0.1.

(\ref{def:rho_c}) 
can be presented in the following form:
\be
x_i^{\min}  = \frac{x_R+x_F}{2-(x_R-x_F)}, \qquad
x_j^{\min}  = \frac{x_i(x_R-x_F)}{2x_i-(x_R+x_F)}~,    
\label{def:xijmin}
\ee
 where $x_R=2p/\sqrt{s}$ and 
     the Feynman variable $x_F$ of the produced hadron, 
     for example the $D$-meson, can be expressed via  
     the variables $p_T$ and $\eta$, or $\theta$ 
     being the hadron scattering angle in the $pp$ c.m.s: 
\begin{eqnarray}
x_F \equiv \frac{2p_{z}}{\sqrt{s}}
=\frac{2p_T}{\sqrt{s}}\frac{1}{\tan\theta}
=\frac{2p_T}{\sqrt{s}}\sinh(\eta). 
\label{def:xFetapt}
\label{def:xFptteta}
\end{eqnarray} 
      One can see from 
(\ref{def:xijmin}) that, at least, one of the low limits $x_i^{\min}$ of the integral 
(\ref{def:rho_c}) must be $\geq x_F$. 
      Thus if $x_F\geq 0.1$, then $x_i^{\min}>0.1$, 
      where the ordinary ({\it extrinsic}) charm distribution is completely negligible
      in comparison with the {\it intrinsic} charm distribution. 
      Therefore, at $x_F\geq 0.1$, or equivalently 
      at the charm momentum fraction $x_c> 0.1$ 
      the {\it intrinsic} charm distribution intensifies the charm 
      PDF contribution into charm hadroproduction substantially
(see Fig.(1)). 
      As a result, 
      the spectrum of the open charm hadroproduction can be increased 
      in a certain region of $p_T$ and $\eta$ (which corresponds 
      to $x_F\geq 0.1$ in accordance to (\ref{def:xijmin})). 
      We stress that this excess (or even the very possibility to observe relevant events in this 
      region) is due to the non-zero contribution of IC component at $x_c > x_F> 0.1$ 
      (where non-IC component completely vanishes).

      This possibility was demonstrated for 
      the $D$-meson production at the LHC in 
\cite{LBPZ:2012}.     
       It was shown that the $p_T$
       spectrum of $D$-mesons is enhanced at pseudorapidities of $3<\eta<5.5$ and 
       10 GeV$/c<p_T<$ 25 GeV/$c$ due to the IC contribution, which was included using the
       CTEQ66c PDF \cite{Nadolsky:2008zw}. For example, due to the IC PDF, with probability about 
       3.5 $\%$, the $p_T$-spectrum increases by a factor of 2 at $\eta=4.5$.
       A similar  effect was predicted in \cite{Kniehl:2012ti}.   

      One expects a similar enhancement in 
      the experimental spectra 
      of the open bottom production 
      due to the (hidden) intrinsic bottom (IB) in the proton, which could have  
      a distribution very similar to the one given in
  (\ref{def:fcPumpl}). 
      However, the probability $w_{\rm IB}$ to find
      the Fock state with the IB contribution $|uudb{\bar b}\rangle$  in the proton 
      is about 10 times smaller than the IC probability $w_{\rm IC}$ 
      due to relation $w_{\rm IB}/w_{\rm IC}\sim m_c^2/m_b^2$ 
\cite{Brodsky:1981,Polyakov:1998rb}.
     
     The IC ``signal'' can be studied not only in the inclusive open (forward) 
     charm hadroproduction at the LHC, but also in some other processes, such as
     production of real prompt photons $\gamma$ or virtual ones $\gamma^*$, or $Z^0$-bosons
     (decaying into dileptons) accompanied by $c$-jets in the kinematics available to the ATLAS and CMS
     experiments.
     The contributions of the heavy quark states in the proton could be 
     investigated also in the $c(b)$-jet 
     production accompanied by the vector bosons $W^\pm,Z^0$. 
     Similar kinematics given by (\ref{def:xijmin}) and (\ref{def:xFptteta}) can also be applied to 
     these hard processes.    

     In the next section we analyze in detail the hard process of the real photon production in 
     $pp$ collision
     at the LHC energies accompanied by the $c$-jet including the IC contribution in the proton.


\subsection{\bf Intrinsic strangeness}
Let us analyze now how the possible existence of the intrinsic strangeness in the proton 
can be visible in $pp$ collisions. For example, consider the $K^-$-meson production in 
the process $pp\rightarrow K^-+X$. Considering the intrinsic strangeness in the proton \cite{IS:2012}
we calculated the inclusive spectrum $ED\sigma/d^3p$
of such mesons within the hard scattering model (Eq.(\ref{def:rho_c})), 
which describes satisfactorily the
HERA and HERMES data on the DIS. The FF and the parton cross sections were taken from
\cite{FF:AKK08,Mangano:2010}, respectively, as mentioned above.
\begin{figure}[h!]
\centerline{\includegraphics[width=0.50\textwidth]{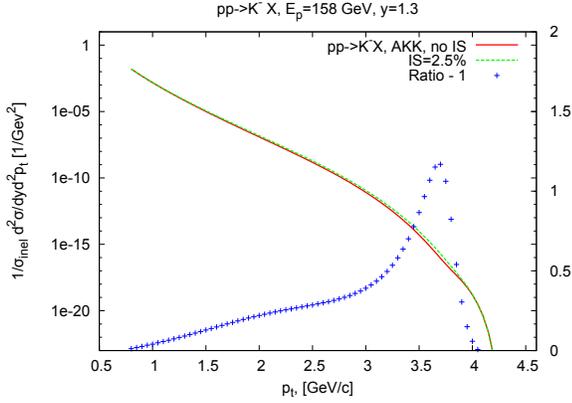}}
 \caption{The $K^-$-meson distributions (with and without intrinsic strangeness contribution) 
  over the transverse momentum $p_t$ for $pp\rightarrow K^- + X$  
  at the initial energy $E=$ 158 GeV, the rapidity $y=$1.3 and $p_t\geq$ 0.8 GeV$/$c
}. 
\label{Fig_5}
\end{figure}
\begin{figure}[h!]
\centerline{\includegraphics[width=0.50\textwidth]{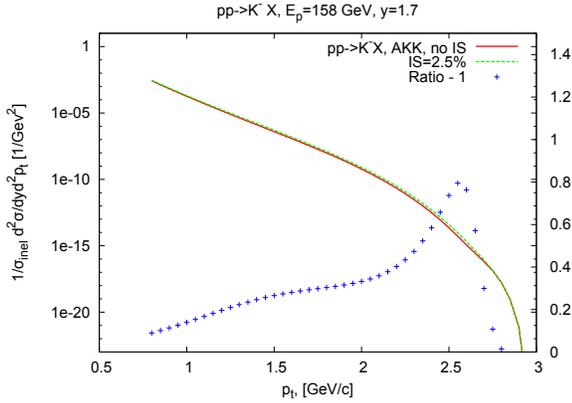}}
 \caption{The $K^-$-meson distributions (with and without intrinsic strangeness contribution) 
  over the transverse momentum $p_t$ for $pp\rightarrow K^- + X$  
  at the initial energy $E=$ 158 GeV, the rapidity $y=$1.7 and $p_t\geq$ 0.8 GeV$/$c
}. 
\label{Fig_6}
\end{figure}
\label{Fig_6}
In Figs.~(\ref{Fig_5},\ref{Fig_6}) the inclusive $p_t$-spectra of $K^-$-mesons produced in $pp$
collision at the initial energy $E_p=$158 GeV are presented at the rapidity $y=$1.3 (Fig~\ref{Fig_5})
and $y=$1.7 (Fig~\ref{Fig_6}). The solid lines in  Figs.~(\ref{Fig_5},\ref{Fig_6}) correspond to our
calculation ignoring the {\it intrinsic strangeness} (IS) in the proton and the dashed curves correspond 
to the calculation including the IS with the probability about 2.5$\%$, according to \cite{IS:2012}. 
The crosses show the ratio of our calculation with the IS and without the IS minus 1.
One can see from Figs.~(\ref{Fig_5},\ref{Fig_6}, right axis) that the IS signal can be above 200 $\%$ 
at $y=$ 1.3, $p_t=$ 3.6-3.7 Gev$/$c and slightly smaller, than 200 $\%$ at 
$y=$ 1.7, $p_t\simeq$ 2.5 Gev$/$c. Actually, this is our prediction for the NA61 experiment that
is now under way at CERN.  

\section{\bf Prompt photon and $c$-jet production} 

    Recently 
    the investigation of prompt photon and $c(b)$-jet production in
    $p{\bar p}$ collisions at $\sqrt{s}=1.96$~TeV was carried out at the TEVATRON 
 \cite{D0:2009}-\cite{Aaltonen:2009wc}.
    In particular,
    it was observed that the ratio of the experimental spectrum of the prompt photons, 
    (accompanied by the $c$-jets) to the relevant theoretical expectation 
    (based on the conventional PDF which ignored the {\it intrinsic} charm) 
    increases with $p_T^\gamma$ up to factor about 3 when $p_T^\gamma$ reaches 110 GeV$/c$.
    Furthermore,  
     taking into account 
     the CTEQ66c PDF,
     which includes the IC contribution obtained within the BHPS model \cite{Brodsky:1980pb,Brodsky:1981} 
     one can reduce this ratio up to 1.5 
\cite{Stavreva:2009vi}. 
     For the $\gamma+b$-jets $p{\bar p}$-production no enhancement 
     in the $p_T^\gamma$-spectrum was observed 
     at the beginning of the experiment
\cite{D0:2009,Aaltonen:2009wc}.
     However in 2012 the D\O\ collaboration has confirmed observation of such an enhancement
\cite{Abazov:2012ea}.   

    This intriguing observation stimulates our interest 
    to look for a similar ``IC signal'' in $p p\rightarrow \gamma+c(b)+X$ 
    processes 
    at LHC energies, see \cite{BDLSS:2014}.

\begin{figure}[h!]
\centerline{\includegraphics[width=0.60\textwidth]{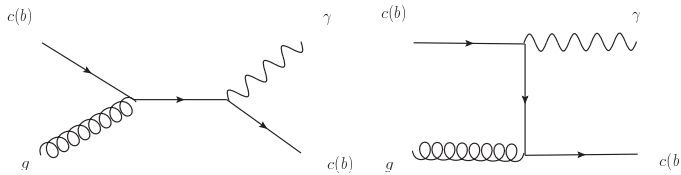}}
 \caption{The Feynman diagrams for the hard process $c(b) g\rightarrow \gamma c(b)$, the one-quark exchange
  in the s-channel (left) and the same in the t-channel (right).}
\label{Fig_Fd_5}
\end{figure}

    The LO QCD Feynman diagrams for the process $c(b)+g\rightarrow\gamma+c(b)$  
    are presented in 
Fig.~\ref{Fig_Fd_5}.  
    These hard sub-processes give the main contribution to the reaction 
    $pp\rightarrow\gamma + c(b)$-jet$+X$.

    Within LO QCD, 
    in addition to the main subprocesses illustrated in 
Fig.~\ref{Fig_Fd_5} 
    one considers the 
    subprocesses $gg\rightarrow c{\bar c}$, $q c\rightarrow q c$, $g c\rightarrow g c$ 
    accompanied by the bremstrallung $c({\bar c})\rightarrow c\gamma$,
    the contribution of which is sizable at low $p_T^\gamma$  and can be neglected at
    $p_T^\gamma>$ 60 GeV$/$c, see \cite{BDLSS:2014} and references therein. 
    The diagrams within the NLO QCD are more complicated than 
Fig.~\ref{Fig_Fd_5}. 

    Let us illustrate qualitatively the kinematical regions where 
    the IC component can contribute significantly 
    to the spectrum of prompt photons produced together with a $c$-jet in $pp$ collisions at the LHC. 
    For simplicity we consider only the contribution 
    to the reaction $pp\rightarrow\gamma+c(jet)+X$
    of the diagrams given in 
Fig.~\ref{Fig_Fd_5}. 
    According to (\ref{def:xFetapt}) and (\ref{def:xijmin}),
   at certain values of the transverse momentum of the photon, $p_T^\gamma$, 
    and its pseudo-rapidity, $\eta_\gamma$, (or rapidity $y_\gamma$)
    the momentum fraction of $\gamma$ 
    can be $x_{F \gamma}>0.1$, 
    therefore the fraction of the initial $c$-quark must 
    also be above 0.1, where the IC contribution in the proton is enhanced 
(see Fig.(1)).
    Therefore, one can expect some non-zero IC signal in the $p_T^\gamma$ 
    spectrum of the reaction $pp\rightarrow\gamma+c+X$ in this
    certain region of $p_T^\gamma$ and $y_\gamma$. 
    In principle, a similar qualitative IC effect can be visible in the production
    of $\gamma^*/Z^0$ decaying into dileptons accompanied by $c$-jets in $pp$ collisions.   

    Experimentally one can measure the prompt photons accompanied by the $c(b)$-jet corresponding
    to the hard subprocess  $c(b) g\rightarrow \gamma c(b)$ presented in Fig.~\ref{Fig_Fd_5}, when 
    $\gamma$ and the $c(b)$-jet are emitted back to back. Therefore, it would be interesting 
    to look at the contribution of this graph to the $p_T^\gamma$ spectrum compared to total QCD calculation 
    including the NLO corrections.    

\begin{figure}[ht]
\centerline{\includegraphics[width=0.60\textwidth]{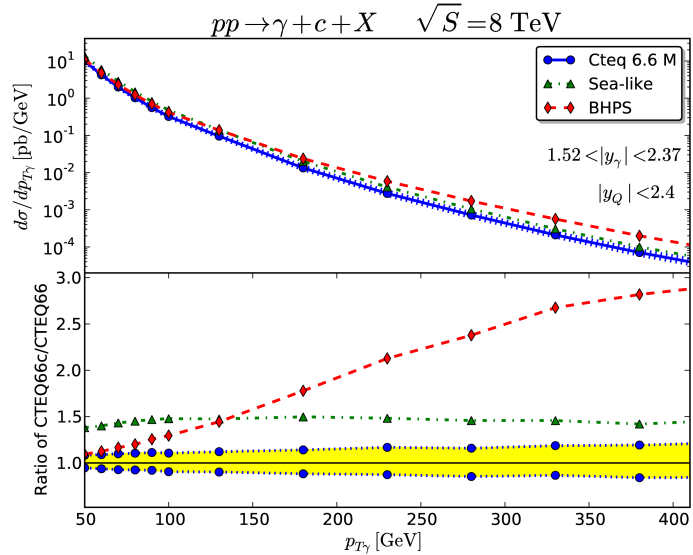}}
\caption{ The $d\sigma/dp_T^{\gamma}$ distribution versus the transverse momentum of the photon for the process $pp\rightarrow\gamma+c+X$
   at $\sqrt{s}=$8 TeV using CTEQ6.6M (solid blue line), BHPS CTEQ6c2 (dashed red line) and sea-like CTEQ6c4 (dash-dotted green line),
   for forward photon rapidity 1.52$<\mid y_\gamma\mid <$2.37.
   The ratio of these spectra with respect to the CTEQ6.6M (solid blue line) distributions (bottom).
   The calculation was done within the NLO QCD approximation.
} 
\label{Fig_Tzvt2}
\end{figure}

  
        In 
Fig.~\ref{Fig_Tzvt2} the differential cross-section $d\sigma/dp_T^\gamma$  calculated at NLO in the massless 
       quark approximation as described in 
\cite{Stavreva:2009vi}  is presented as a function of the transverse momentum of the prompt photon.
       The following cuts are applied: 
       $p_T^\gamma>45$~GeV,  $p_T^c>20$~GeV 
        with the $c$-jet pseudorapidity in the interval $\mid y_c\mid\leq 2.4$ and the photon 
        pseudorapidity in the  forward region of  photon rapidities $1.52<\mid y_\gamma\mid<2.37$. 
        The solid blue line represents the differential cross-section calculated with the 
        radiatively generated charm PDF (CTEQ66), the dash-dotted green line uses as input 
        the sea-like PDF (CTEQ66c4) 	and the dashed red line the BHPS PDF (CTEQ66c2).
        In the lower half of Fig.~\ref{Fig_Tzvt2} the above distributions normalized to the distribution 
        acquired using the CTEQ66 PDF  and $\mu_r=\mu_f=\mu_F=p_T^\gamma$, are presented.  
        The shaded yellow region, represents the scale dependence.  
        Clearly the difference between the spectrum using the BHPS IC PDF and the one using the radiatively generated PDF 
        increases as $p_T^\gamma$ increases.

	Therefore Fig.~(\ref{Fig_Tzvt2}) shows that the 
	IC signal could be visible at the LHC energies with both the ATLAS and CMS detectors 
	in the process $pp\rightarrow\gamma+c+X$ when $p_T^\gamma\simeq $ 150 GeV$/$c. 
	In the region the IC signal dominates over the all non-intrinsic charm background 
	with significance at a level of a factor of 2 (in fact 170\%).     

\section{$W/Z$-boson and $b,c$-jet production}

\begin{figure}[h!]
\centerline{\includegraphics[width=0.80\textwidth]{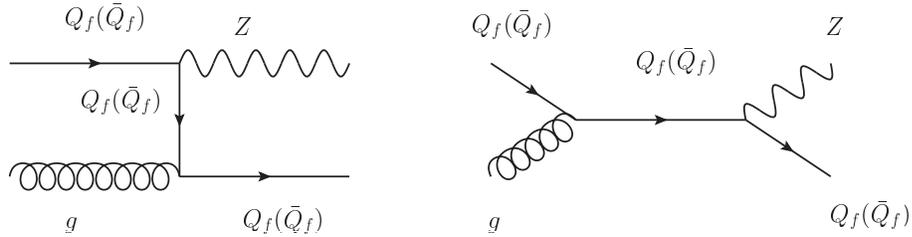}}
\caption{LO Feynman diagrams for the process $Q_f({\bar Q}_f) g\rightarrow Z Q_f({\bar Q}_f)$, where 
$Q_f=c,b$
}.
\label{Fig_Qg_ZQ}
\end{figure}

Let us analyse now another process, the production of vector boson accomponied by the $b$-jet 
in $pp$ collision. The LO QCD diagram for the process $Q{\bar Q}_f)+g\rightarrow Z^0 +Q_f({\bar Q}_f)$ 
is presented in Fig.~\ref{Fig_Qg_ZQ}. These hard subprocesses can give the main contribution to the 
reaction $pp\rightarrow Z^0(\rightarrow l^+ + l^-)+Q_f({\bar Q}_f)-jet +X$, which could give us also 
the information on the {\it IC} contribution in the proton.  
\begin{figure}[h!]
\centerline{\includegraphics[width=0.40\textwidth]{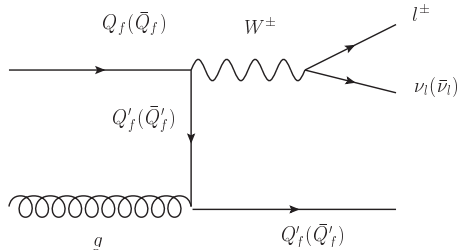}}
\caption{Example of an LO Feynman diagram for the process $Q_f({\bar Q}_f) g\rightarrow W^\pm Q_f^\prime({\bar Q}_f^\prime)$,
                where $Q_f=c.b$ and $Q_f^\prime=b,c$ respectively.
}.
\label{Fig_Qg_WQ}
\end{figure}
 The LO QCD diagram for the process $Q{\bar Q}_f)+g\rightarrow W^{\pm} +Q^\prime_f({\bar Q}^\prime_f)$ 
is presented in Fig.~\ref{Fig_Qg_WQ}, where $Q_f=c,b$ and  $Q^\prime_f=b,c$. These hard subprocesses 
can give the main contribution to the reaction $pp\rightarrow W^\pm(\rightarrow l^+ + \nu)+Q^\prime_f({\bar Q}^\prime_f)-jet +X$, 
which could give us the information not only on the {\it IC} contribution but also on the {\it IS} one in the proton.  

\begin{figure}[h!]
\centerline{\includegraphics[width=0.40\textwidth]{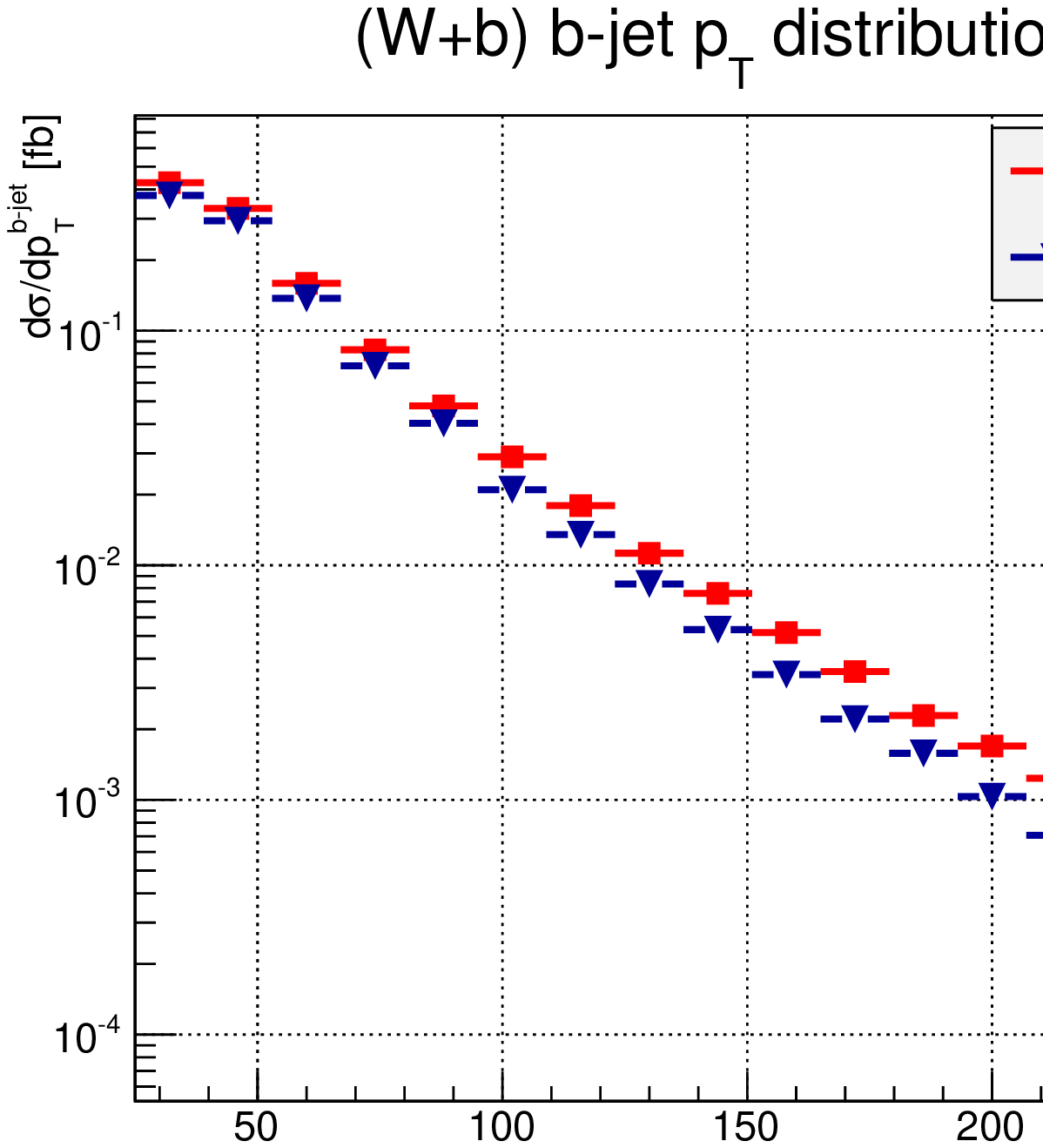}}
\centerline{\includegraphics[width=0.40\textwidth]{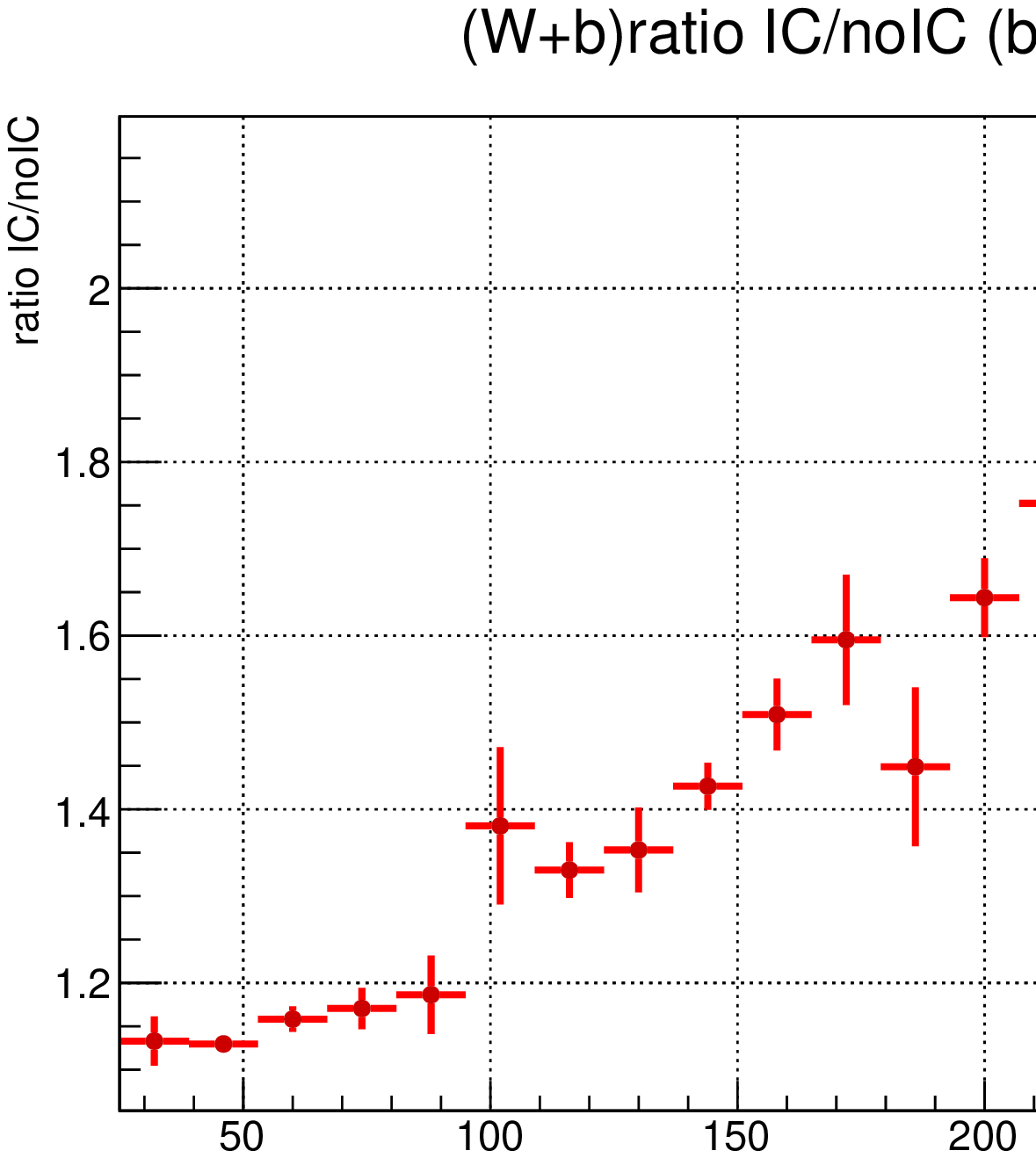}}
 \caption{ Comparison of the $p_T$-spectra for the NLO $pp\rightarrow W+b$,
               processes 12+17 \cite{MCFM} obtained with
               PDF including intrinsic charm component (CTEQ66c) and PDF having only extrinsic component 
              (CTEQ66, top). Bottom: Ratio of these two spectra.
}. 
\label{Fig_Wb9}
\end{figure}

\begin{figure}[h!]
\centerline{\includegraphics[width=0.40\textwidth]{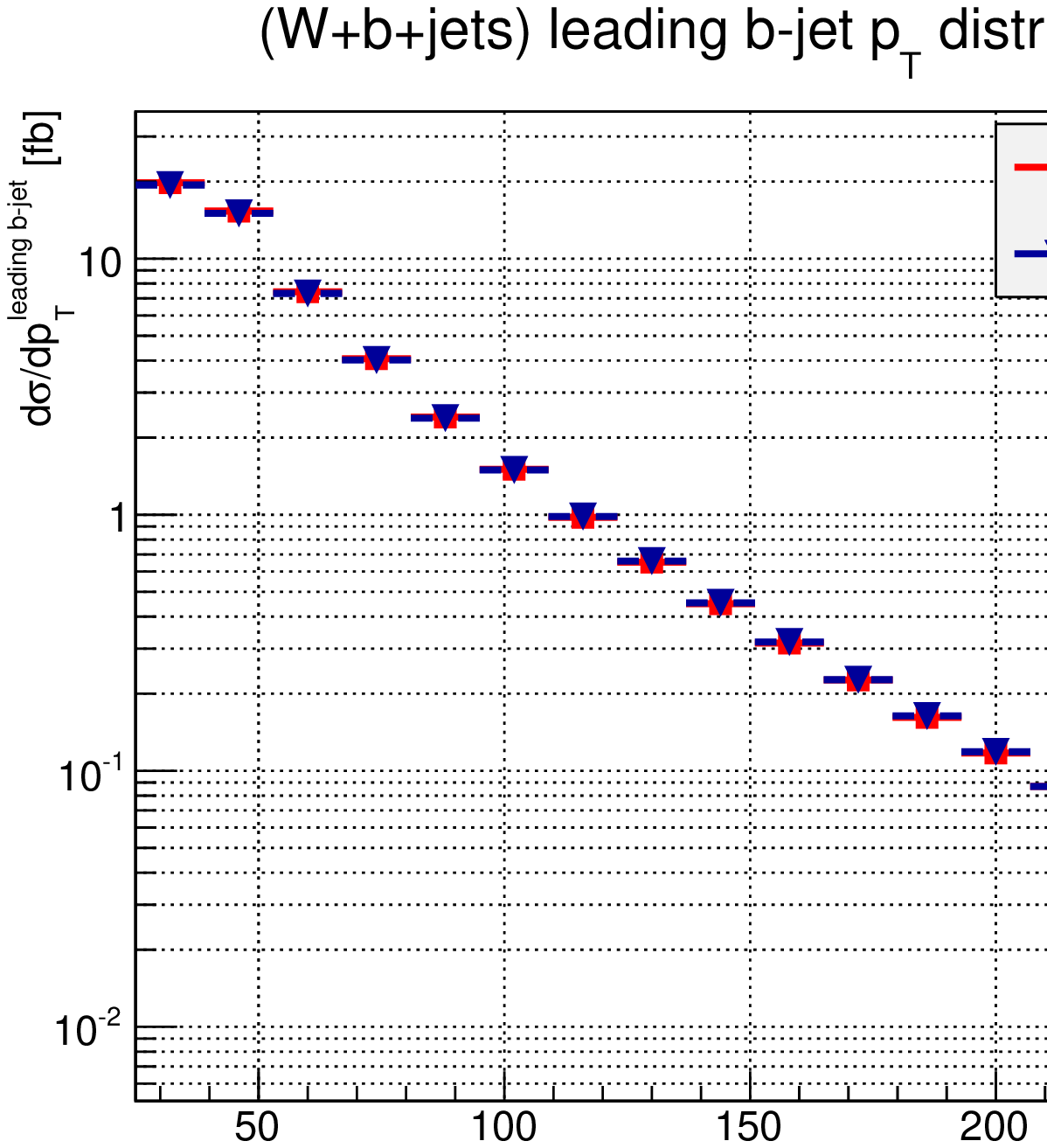}}
\centerline{\includegraphics[width=0.40\textwidth]{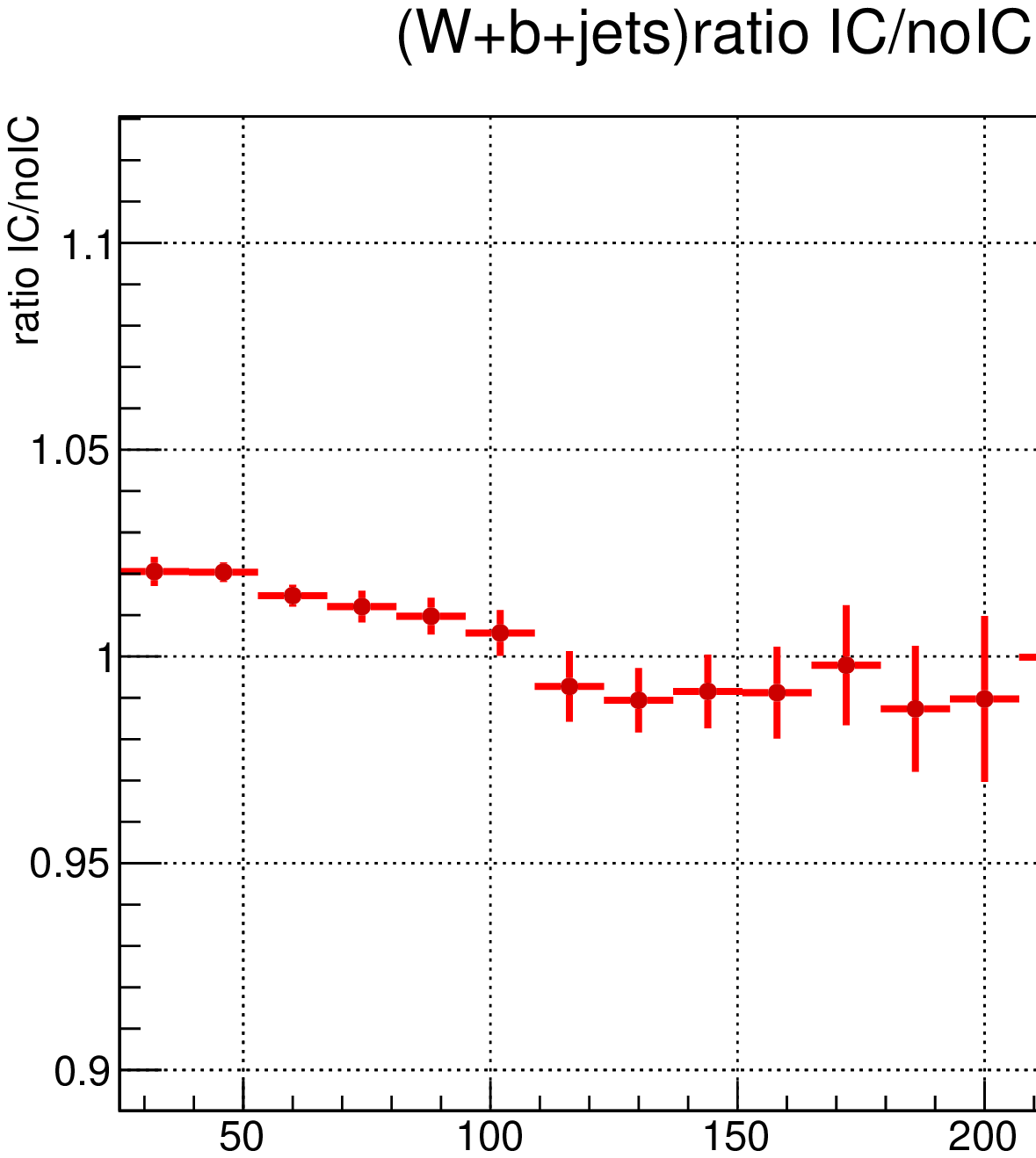}}
\caption{ Comparison of the $p_T$-spectra for the  NLO $pp\rightarrow W+b +jet$ processes 
               401/406 and 402/407 \cite{MCFM} obtained with
               PDF including intrinsic charm component (CTEQ66c) and PDF having only extrinsic component 
               (CTEQ66, top). Bottom: Ratio of these two spectra.  
}. 
\label{Fig_Wb10}
\end{figure}

\begin{figure}[h!]
\centerline{\includegraphics[width=0.40\textwidth]{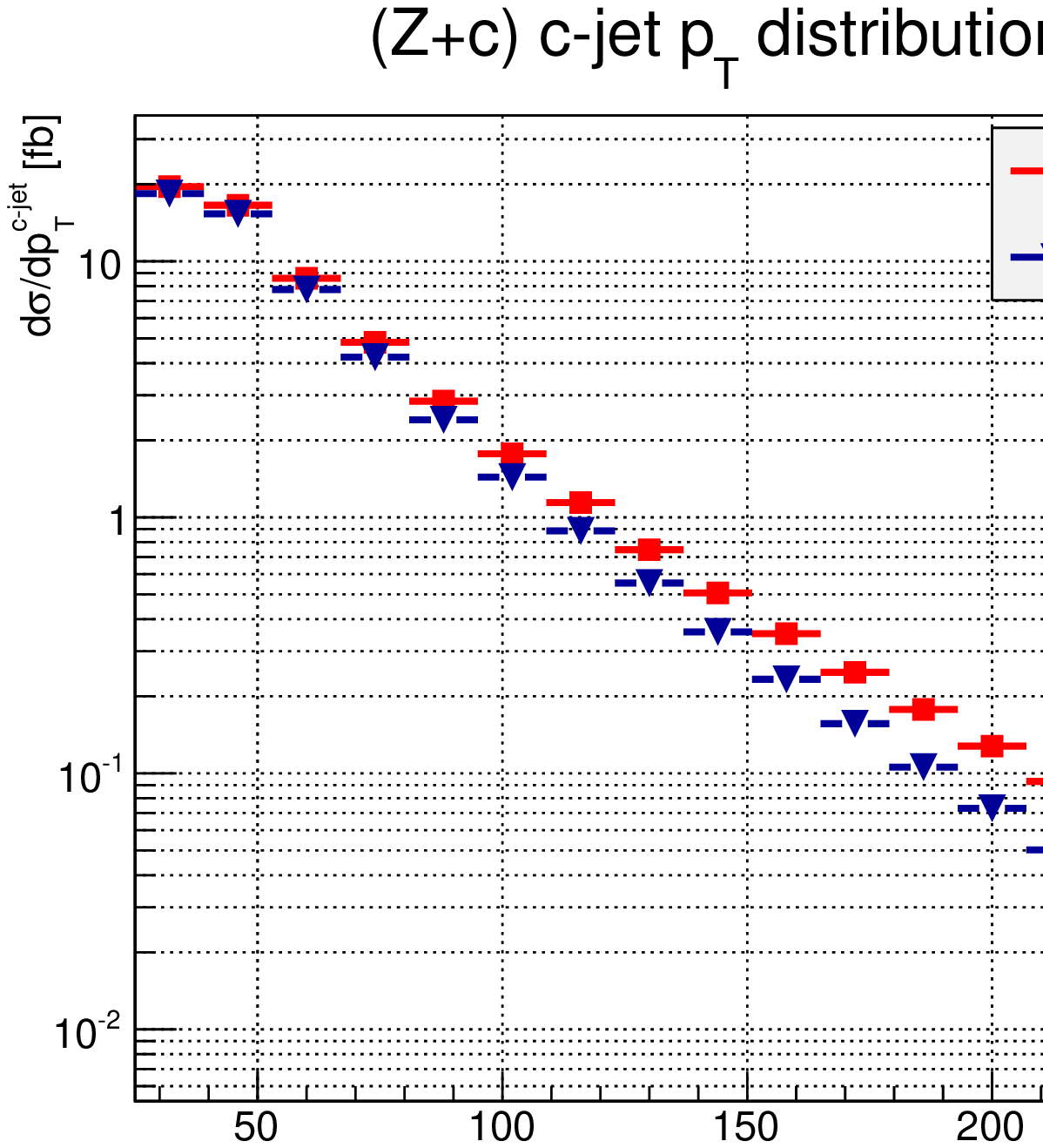}}
\centerline{\includegraphics[width=0.40\textwidth]{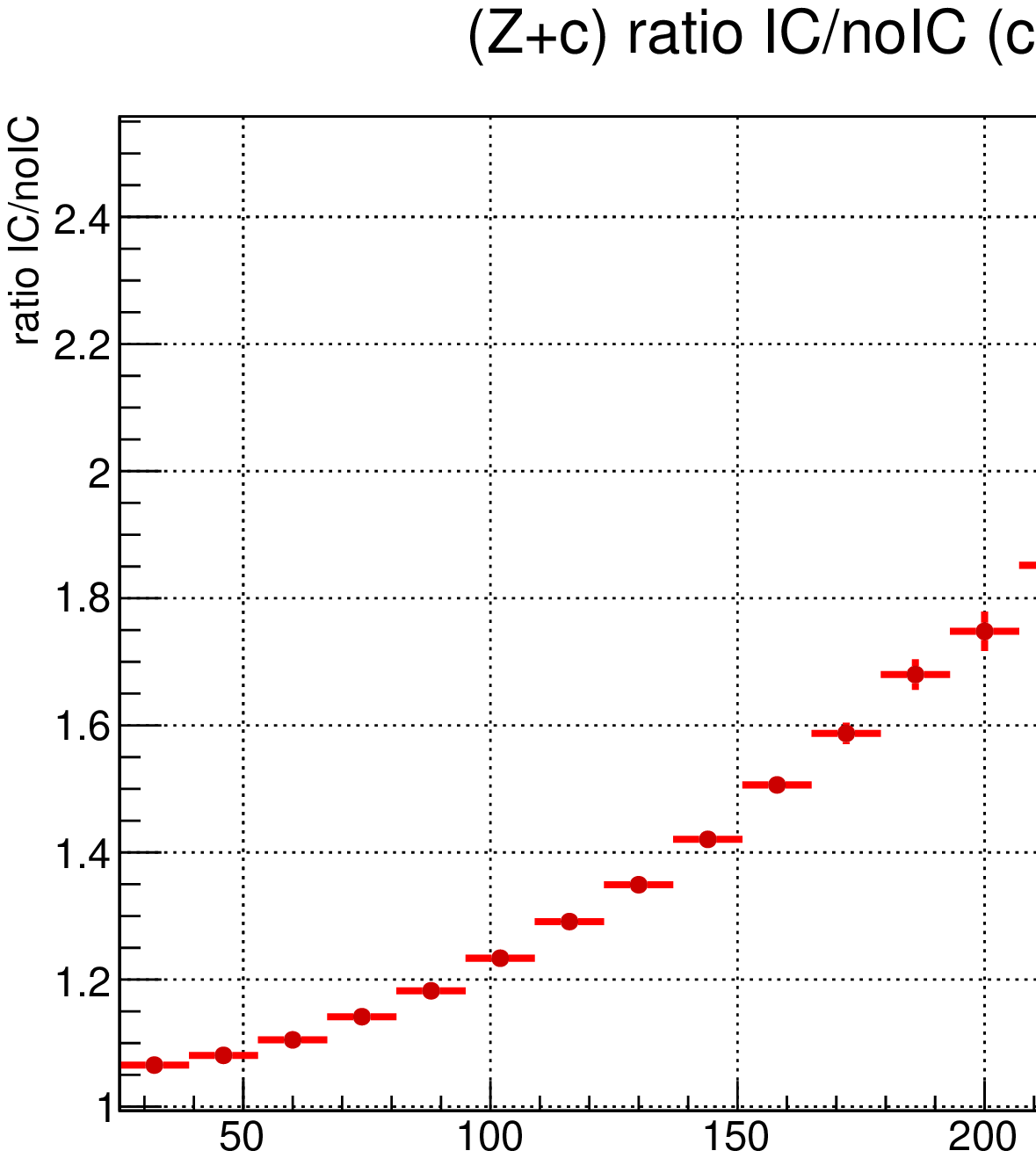}}
 \caption{Comparison of the $p_T$-spectra for the total NLO $pp\rightarrow Z+c$ {\bf process 262} obtained
               with PDF including intrinsic charm component (CTEQ66c) and PDF having only extrinsic component 
	       (CTEQ66, top). Bottom: Ratio of these two spectra.
}. 
\label{Fig_Zc11}
\end{figure}

\begin{figure}[h!]
\centerline{\includegraphics[width=0.40\textwidth]{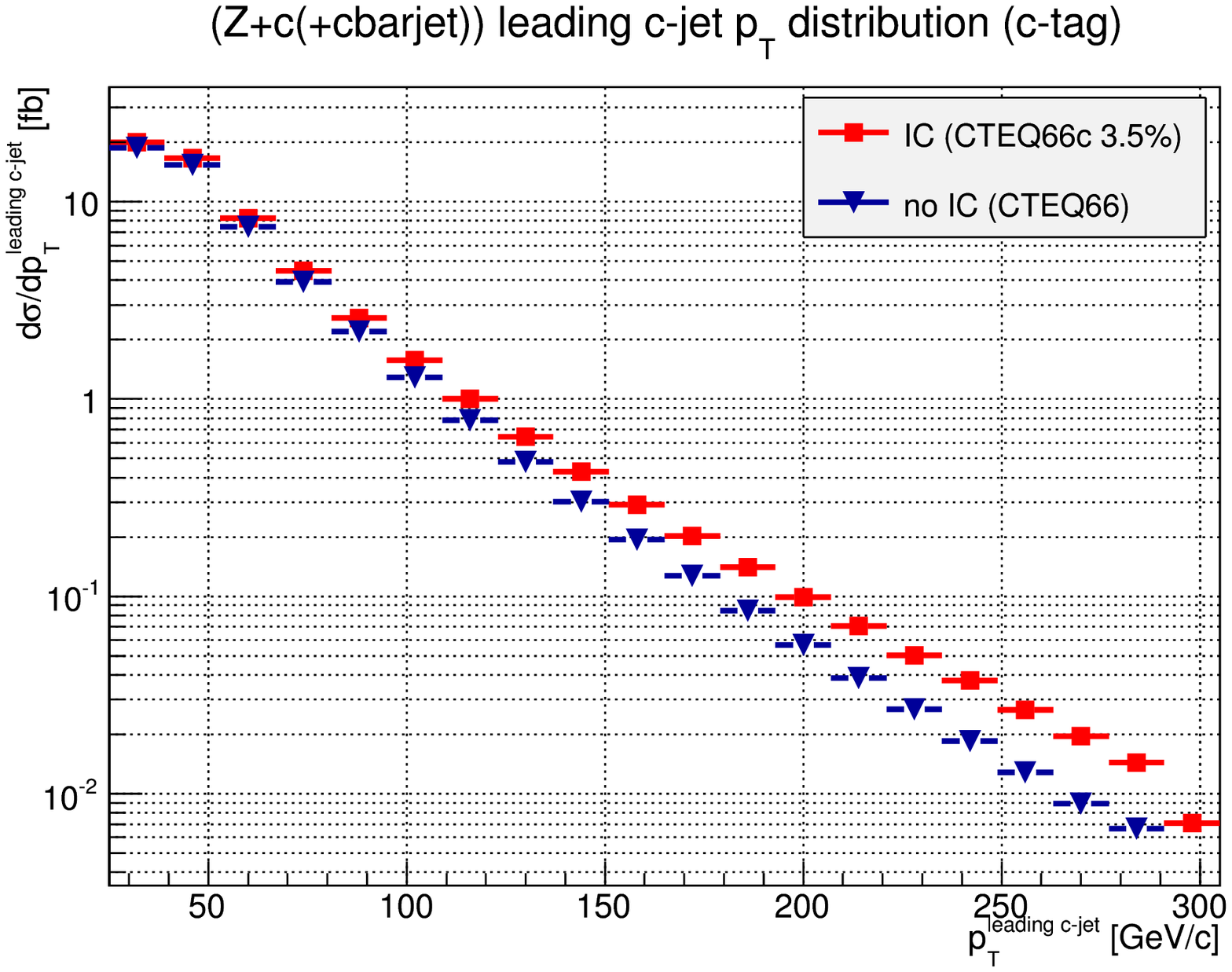}}
\centerline{\includegraphics[width=0.40\textwidth]{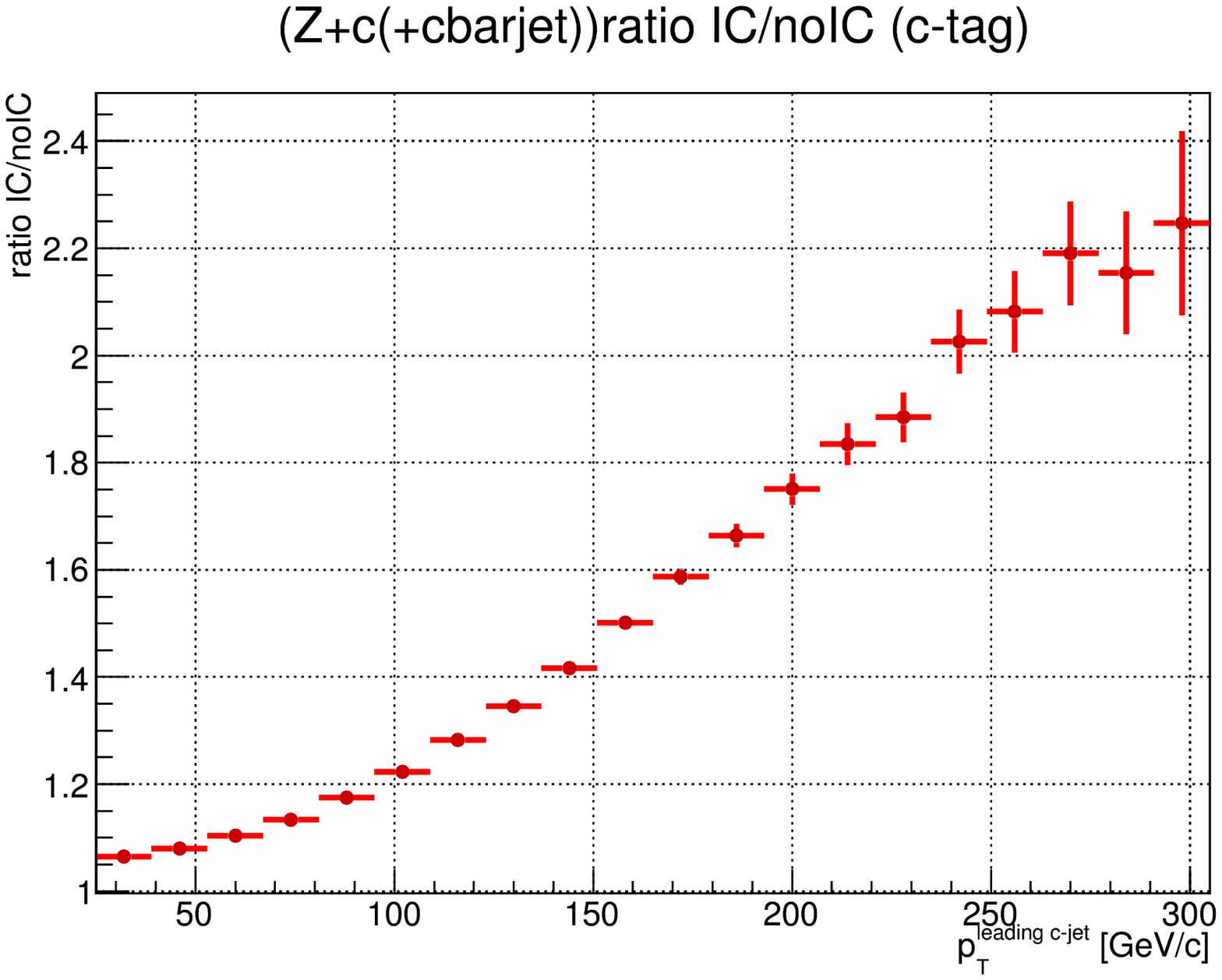}}
 \caption{ Comparison of the $p_T$-spectra for the total NLO 
               processes  $pp\rightarrow Z+c(\bar{c})$ ( processes 262,264,267 \cite{MCFM} )
               obtained with PDF including intrinsic charm component 
	       (CTEQ66c) and PDF having only an extrinsic component (CTEQ66). Heavy flavor jet
	       tagging efficiencies have been applied to the $c$-jets and the $b$-jets (top). 
               Bottom: Ratio of these two spectra.
}. 
\label{Fig_Zc12}
\end{figure}

In Fig.~\ref{Fig_Wb9} (top) the transverse momentum spectrum of $W^+$-boson accomponied by the $b$-jet
produced in $pp$ collision at the LHC energy $\sqrt{s}=$ 8 TeV is presentedc calculated within the MCFM generator,
see processes 12+17 \cite{MCFM}. 
The down line corresponds to the calculation without the IC, the upper curve is our result including the IC contribution 
in the PDF CTEQ6,6c withthe probability about 3.5 $\%$.  
In Fig.~\ref{Fig_Wb9} (bottom) the ratio of spectra with and without {\it IC} contribution the proton PDF 
is presented. It is seen that the {\it IC} contribution grows when the transverse momentum of the $b$-jet
$p_T^{b jet}$ increases and becomes a factor about 2 at $p_T^{b jet}=$ 250-300 GeV$/$c. However, the cross section
of this proccess is too low especially alt large $p_T^{b jet}$. Therefore, we calculated within the NLO the
distribution of the leading $b$-jet produced in $pp$ collision at $\sqrt{s}=$ 8 TeV in association with $W^\pm$-boson and
another jets (heavy anf light) as a function of its transverse momentum $p_T^{leading b jet}$, proceses 401+406+402+407
\cite{MCFM}. 
This spectrum is presented in 
Fig.~\ref{Fig_Wb10} (top) with {\it IC} and without {\it IC} contribution in PDF. It is seen that the cross section of these
processes is much larger the one for processes presented in Fig.~\ref{Fig_Wb10}. However, in these processes there is no the
{\it IC} signal at any 50 GeV$/$c$<p_T^{leading b jet}<$300 GeV$/$, as is seen from Fig.~\ref{Fig_Wb10} (bottom). 
Therefore, we calculated within the NLO the similar $p_T$-spectrum of the $c$-jet produced in $pp$ collision at 
$\sqrt{s}=$ 8 TeV in association with $Z^0$-boson using the PDF with and without the {\it IC} contribution. This spectrum
is presented in Fig.~\ref{Fig_Zc11} (top). It is seen that the cross section is the same order as the cros section for
$pp\rightarrow W+b +jet$ presented in Fig.~\ref{Fig_Wb10}. The {\it IC} signal in this process grows when $p^{c jrt}_T$
increases and becomes a factor about 2.4 at $p_T^{c jet}=$ 250-300 GeV$/$c, as is seen from Fig.~\ref{Fig_Zc11} (bottom). 
In Fig.~\ref{Fig_Zc12} (top) the spectrum of the leading $c$-jet produced in $pp$ collision at$\sqrt{s}=$ 8 TeV in 
association with $Z^0$-boson and other $c$- and light jets is presented ( processes 262,264,267 \cite{MCFM} . We used 
the PDF with and without the {\it IC} contribution and the $c$-jet tagging efficiency has been applied.   

\begin{figure}[h!]
\centerline{\includegraphics[width=0.40\textwidth]{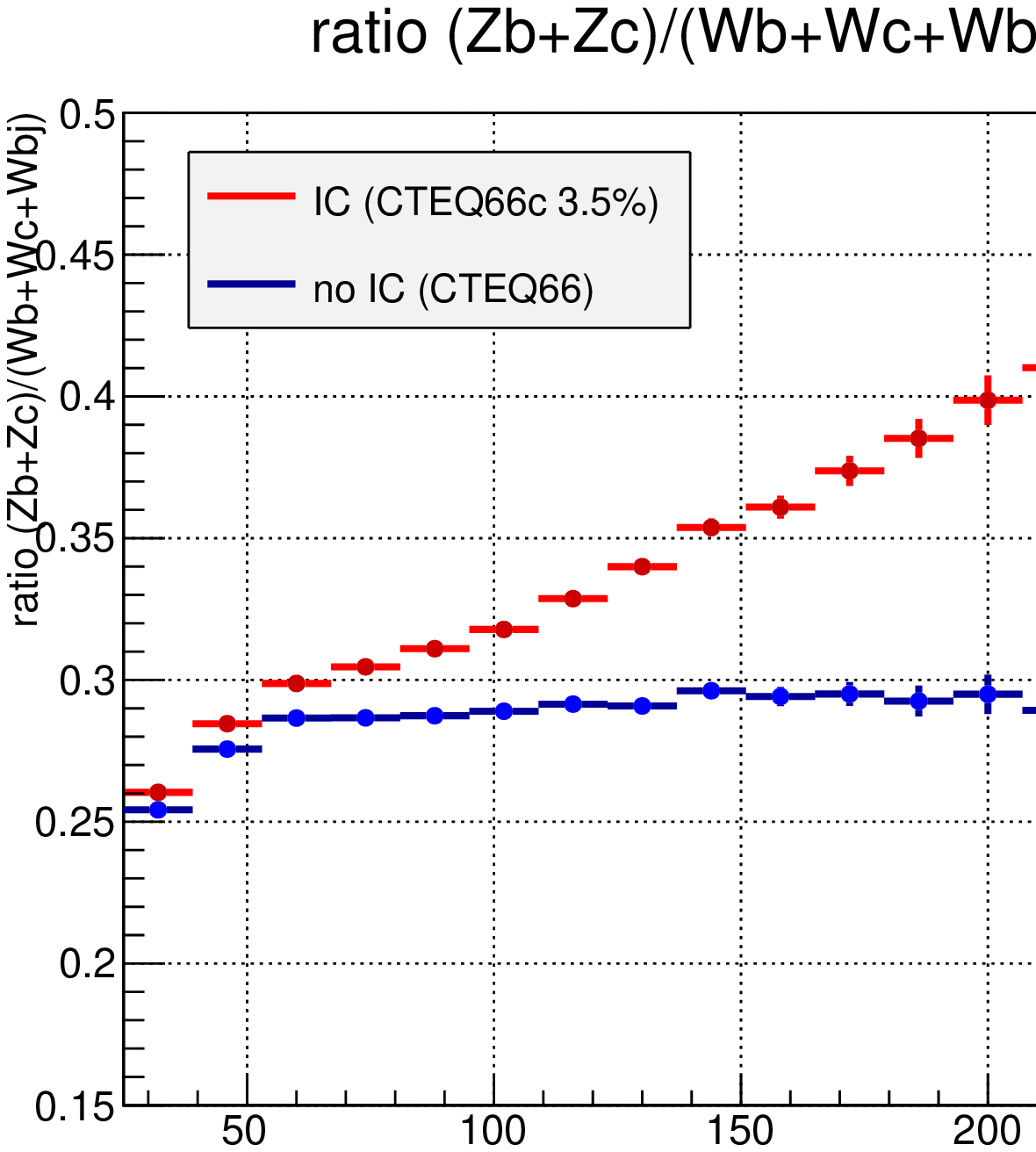}}
\centerline{\includegraphics[width=0.40\textwidth]{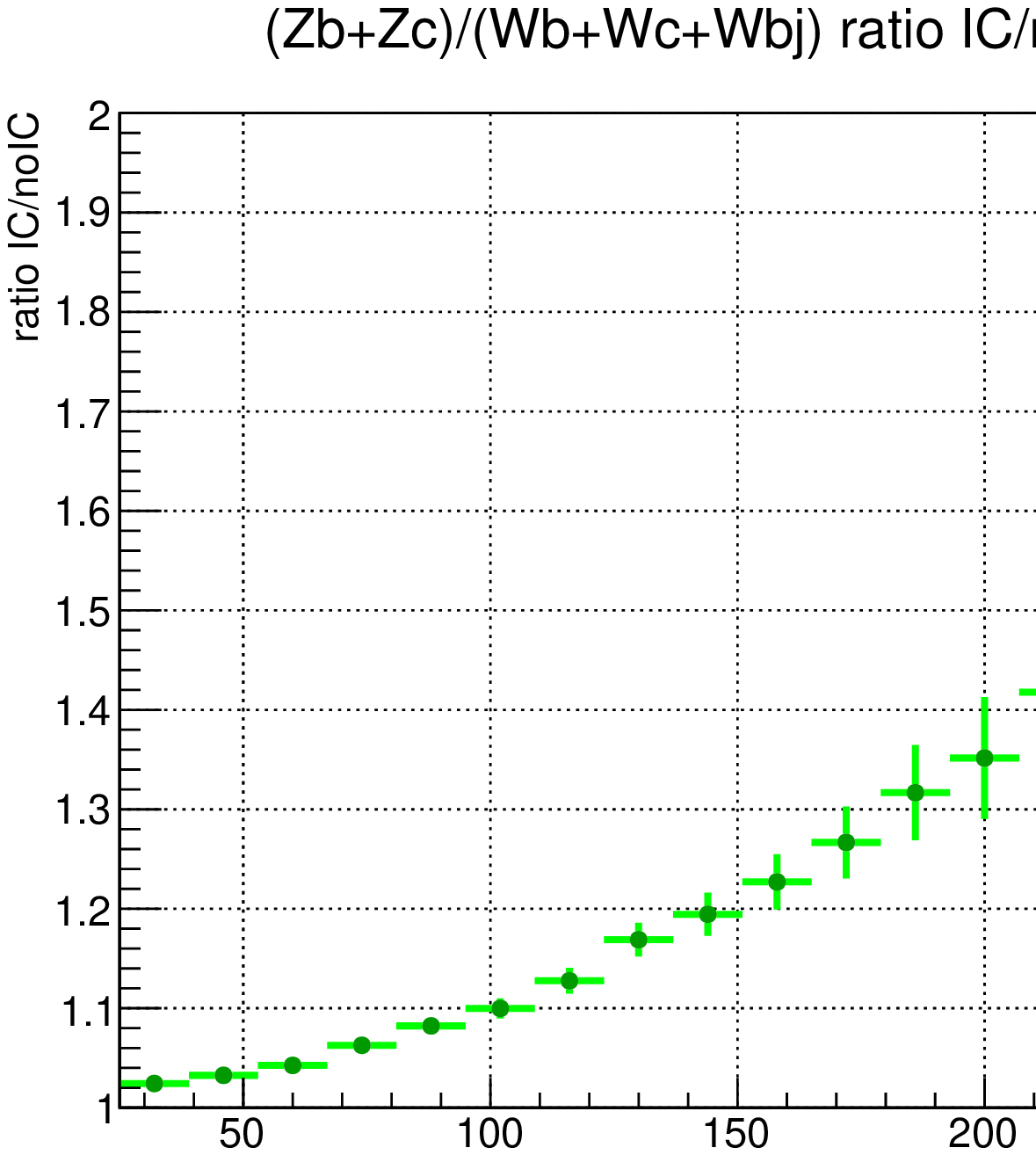}}
 \caption{Comparison of the ratio of the $p_T$-spectra for $Z+Q$ to $W+Q$  processes obtained with 
	      PDF including intrinsic charm component (CTEQ66c) and PDF having only an extrinsic 
	      component (CTEQ66). Heavy flavour jet tagging efficiencies have been applied to the $c$-jets 
	      and the $b$-jets (top). Bottom:  Ratio of these two ratios of spectra
}. 
\label{Fig_ZW_cb13}
\end{figure}

\newpage
\section{Conclusion} 

          We analyzed the inclusive $K^-$-meson production in $pp$ collision at the initial energy 
          $E_p=$158 GeV and gave some predictions for the NA61 experiment going on at CERN. We showed that
          in the inclusive spectrum of $K^-$-mesons as a function of $p_t$ 
          at some values of their rapidities
          the signal of the {\it intrinsic} strangeness can be visible and reach about 200$\%$ and more at large 
          momentum transfer we took.      
          The probability of the {\it intrinsic} strangeness to be about 2.5$\%$, as 
          was found from the best description of the HERA and HERMES data on the DIS, see \cite{IS:2012} and
          references therein.
          The similar predictions can be made for the open strangeness production at the energies of the 
          CBM (Darmstadt) and NICA (Dubna) experiments. The main goal of such predictions is to show that 
          at the certain kinematical region the contribution of the {\it intrinsic} strangeness in the proton
          can result in the enhancement in inclusive spectra by a factor of 2-3. This enhanced strangeness in the 
          nucleon can lead to the enhanced yield of the strange hadrons produced in $AA$ collisions at NICA and 
          CBM in the kinematical region forbidden for the free $NN$ collisions. 

      We have shown also that the possible existence of an intrinsic heavy quark component 
      to the proton can be seen not only in the forward open heavy flavor production of $pp$-collisions
      (as it was believed before) but it can also be observed in the semi-inclusive $pp$-production of 
      massive vector bosons in association with heavy flavor jets ($b$, and $c$). In particular, 
      it was shown that the {\it IC} contribution can produce much more $Z+c$-jet events (factor 1.5 -- 2) than
      what is predicted from extrinsic contribution to PDF alone, when the heavy flavor jet has a transverse
      momentum of $p^{jet}_{T}>$ 100 GeV$/$c and a pseudo-rapidity satisfying 1.5$<\mid \eta_{jet}\mid<$ 2.4.
      We then showed that this conclusion stays true when the $Z+b$ negative contribution and the inefficiencies in 
      the experimental identification of heavy flavor jets are taken into account. 
     
      We then showed that because of the dominant contribution of gluon-splitting processes, the production 
      of W-bosons accompanied by heavy flavor jets is not sensitive to intrinsic quarks. We took advantage
      of this to propose a promising measurement that reduces the expected systematic uncertainties on the 
      measurement results compared to a differential cross section measurement of the heavy flavor jet spectrum in 
      $Z+Q$ events. The idea is to use the ratio of the leading heavy flavor spectra in inclusive heavy flavor
      $Z+Q$ to $W+Q$ events to verify the predictions about an {\it IC} contribution to the proton. 
      Such measurements can already be made with ATLAS and CMS available data.

\section{Acknowledgements}
We thank S.J.Brodsky,  M. Gazdzicki, S.M. Pulawski and  A.Rustamov
for extremely helpful discussions and recommendations for the 
predictions on the search for the possible intrinsic heavy flavour components in
$pp$ collisions at high energies.
We are also grateful to P-H.Beauchmin for very productive collaboration, H.Jung, A.Likhoded, A.V.Lipatov, 
and N.P.Zotov for very helpful discussions. 
This work was supported in part by the Russian Foundation for Basic Research, 
grant No: 13-02001060 (Lykasov).

\begin{footnotesize}

\end{footnotesize}

\end{document}